# Students’ Validation Criteria for AI-Generated Physics Solutions

*Accepted for publication in The Physics Teacher*

Álvaro Suárez[1], Richard González[2]
[1,2]Consejo de Formación en Educación, Instituto de Profesores Artigas, Montevideo, Uruguay
[1]alsua@outlook.com, [2]rgonzalez@niecyt.exa.unicen.edu.ar

A physics solution can be clear without being valid. This distinction becomes important when students read AI-generated solutions that appear coherent but still require physical and mathematical checking. We examined how pre-service physics teachers in a calculus-based introductory electromagnetism course evaluated AI-generated solutions to a charged-rod electric-field problem after attempting the problem themselves. Although the AI-generated solutions analyzed were correct, no student accepted a response as valid after a first reading. Students who had solved the problem fully or in part could examine the responses using physical and mathematical criteria, while others withheld acceptance because they could not follow the mathematical development. These results suggest that, in physics instruction, AI-generated solutions should be treated less as answers to be accepted or rejected and more as worked-out responses that students learn to reconstruct, question, and test.

## 1. Introduction

In recent years, generative AI has become increasingly visible in physics teaching and learning. Recent work has examined both its potential to support instruction and the challenges raised when students use these tools without much critical scrutiny, as well as the roles AI can play in physics teaching—as a tutor, a tool, or an object of analysis.[1-4]
In this context, a particularly important question is what happens when a student reads an AI-generated solution and must decide whether it is valid. Rather than asking only whether the answer is correct, this article examines the criteria students use to judge the validity of that solution.

This question matters instructionally because using AI critically requires more than warning students that AI can be wrong. The key issue is understanding what features of a solution lead students to accept it as valid or to withhold acceptance until they can examine it further. These judgments can also provide evidence about whether students attribute authority to the AI tool itself. The need to examine students’ validation criteria becomes especially relevant when tools can produce polished, plausible responses that appear reliable before students have examined the reasoning behind them. In the sections that follow, we describe the classroom activity, analyze students’ validation judgments, and discuss implications for teaching.

## 2. The activity

The activity was implemented during a two-and-a-half-hour class with 20 pre-service physics teachers in a calculus-based introductory electromagnetism course. The full activity is available in Appendix A. The sequence began with an approximately 15-minute individual phase before students were given the physics problem or AI was mentioned. Students were asked to imagine a classmate’s solution and write down three features of the solution they would examine when deciding whether it was valid. This phase made their initial validation criteria visible before they encountered the specific problem.

In the second phase, students worked in small groups for approximately 75 minutes on a problem about the electric field of a uniformly charged rod. The task was to determine the field components at a point on the line perpendicular to the rod passing through one end and to analyze two limiting cases: one in which the distance to the observation point was much larger than the length of the rod, and the opposite limit, in which the rod was much longer than that distance. To solve the problem, students had to make decisions about representation, modeling, physical interpretation, and the use of integral calculus. The choice of a problem requiring integral calculus was deliberate, as the subsequent validation task required students to follow a nontrivial mathematical development rather than judge only whether a solution sounded coherent.

The third phase introduced AI and lasted approximately 30 minutes. Each student asked an AI tool, such as ChatGPT or Gemini, to solve the problem. Students were free to choose both the tool and the prompt, so that this phase would resemble the way they normally use these systems outside the activity. They then read the full response once. Before any group discussion or further interaction with the AI tool, students chose one of three options: accept the solution as valid after that first reading; lean toward considering it valid but still need to check one or two specific aspects; or not accept it as valid. Choosing the last option did not necessarily mean judging the solution to be wrong. This sequencing focused attention on students' first judgment of a specific solution. Students then explained what they had looked at in making that decision and, when applicable, what they would check before accepting the solution.

Finally, in an approximately 30-minute phase, students returned to their original small groups to compare their individual judgments with one another and with the criteria they had written at the beginning of the activity and to discuss what the activity had helped them notice about evaluating an AI-generated solution. A brief whole-class exchange concluded the phase.

Student responses were analyzed jointly by both researchers. After an initial joint review of the full set of responses, preliminary categories were developed inductively and refined as the responses were classified.

## 3. Results

### *3.1 Initial criteria stated by students*

Before working on the specific problem and with no reference to AI, students tended to state mostly physics-based and procedural criteria for judging a solution. Their responses focused on five areas: alignment with the problem statement; analysis of the diagram or representation; correctness of the mathematical procedures; identification of given information and assumptions; and the physical reasonableness of the result.

One student summarized this pattern by answering that they checked "what kind of phenomenon is involved, what quantities are involved, and which laws are being used," whether the equations were "appropriate and correctly applied," and whether the result made physical sense, including its units and sign. Overall, students mainly drew on physics content and how the reasoning unfolded.

### *3.2 Solving the problem*

Eight of the 20 students solved the problem fully or in part, while the remaining 12 did not reach a consistent result. The main difficulty among these 12 students was setting up and evaluating the integrals required to construct the field by superposition.

### *3.3 Student interaction with AI: prompts and first validation judgments*

Among the 20 students, 17 used ChatGPT (GPT-5.4 mini), while three used Gemini (Gemini 3 Flash). Students' prompts varied in specificity. Ten students gave short, general instructions, such as "Solve this problem" or “How would you solve it?” Seven students added context, asking for a step-by-step solution, specifying the course level, or requesting explanations. Only three students used more elaborate prompts, asking the AI to follow a textbook-like approach, state assumptions explicitly, and explain the physical reasoning behind each step. All 20 AI-generated solutions were reviewed by both researchers through the conversation links provided by the students and were found to be physically and mathematically correct in their reasoning and results. The solutions did not go beyond the physics and mathematics expected of students in the course, although they varied in level of detail (see Appendix B for one complete example from each model).

After reading the AI-generated solution, no student accepted it as valid after that first reading, as shown in Table 1. Sixteen students leaned toward considering it valid but still needed to check one or two specific aspects, while four did not accept it as valid. We found no relationship between prompt specificity and students’ first validation judgments.

Their explanations show that the initial evaluation did not rest on a single criterion. Some students were cautious because the problem was complex and a small error in an early step could affect the rest of the solution. Others focused on whether the response made enough of its reasoning explicit for them to examine it. As one student wrote, “not because I think it is wrong, but because there are steps it skips that seem important to me.” This concern with missing or insufficiently explained steps was especially visible among students who had not reached a consistent result on their own. For this group of 12 students, the issue was mathematical: they had difficulty following the mathematical development in the AI-generated solution. This did not always lead them to the same initial judgment: two students did not yet accept the response as valid, while the remaining ten leaned toward considering it valid but asked for specific checks. The main difficulty was not identifying a specific error in the AI-generated result but tracing the intermediate steps it provided. One student wrote, “Since I do not understand the integration steps, I cannot accept the solution as valid because I do not know whether they are correct.” Another said they would need to ask the AI to explain the integrals and substitutions “step by step” until they could follow the reasoning from what they already knew.

Table 1. Students’ problem-solving outcome and first validation judgment of the AI-generated solution

| Problem-solving outcome \ First validation judgment | Accepted as valid | Leaned toward valid, checks needed | Did not accept | Total |
|---|---|---|---|---|
| Solved fully or in part | 0 | 6 | 2 | 8 |
| Did not reach a consistent result | 0 | 10 | 2 | 12 |
| Total | 0 | 16 | 4 | 20 |

Students who had solved the problem fully or in part evaluated the AI-generated solution by examining the physics and mathematics of the response. Some compared it with their own work or with the approach usually used in the course. One student did not yet accept the solution as valid because the AI used “a vector approach” and “that is not the way we are solving it in class,” adding that “it needs more explanation” to understand the connection between the calculation used in class and the vector approach. Others used criteria like those they had stated in the first phase of the activity. In these cases, students focused on the correct use of signs, quantities, and units; the logic of the procedure; consistency with the problem statement; and checking the given information and assumptions. One student wrote, “I mainly looked at whether the signs were used correctly, whether the quantities and their units were correctly interpreted and applied, and whether the procedure followed a logic consistent with the problem statement."

## 4. Discussion and implications for teaching

The results of this activity show that students' evaluation of an AI-generated solution was not simply a matter of accepting it as valid or withholding acceptance after a first reading. Students who had solved the problem fully or in part withheld acceptance while examining the response using physical and mathematical criteria related to the problem statement, the representation, the solution process, and the reasonableness of the result. Students who struggled with the problem also withheld acceptance, mainly because they could not follow the mathematical procedures. Thus, caution corresponded to different processes: in some cases, disciplinary review; in others, limited access to the reasoning needed to judge the solution.

These validation judgments provide a limited but useful window into the authority that students appeared to attribute to AI. Students did not base their judgment on general trust or distrust in AI; they treated validity as a judgment about a particular solution. Previous studies have raised concerns that students may accept incorrect ChatGPT responses as correct[1] or that high linguistic quality may make scientifically limited AI-generated responses appear more scientifically accurate, especially when students have limited prior knowledge.[5] In the activity analyzed here, however, after first stating validation criteria and attempting the problem themselves, no student accepted the response as valid after a first reading. This criterion-based approach is consistent with a study involving prospective physics teachers in which structured criteria helped participants evaluate ChatGPT-generated content more critically and recognize its limitations.[6]

Students' interactions with AI varied in how specific their prompts were, but this variation was not reflected in their first validation judgments. This should be interpreted cautiously: it does not mean that prompting is unimportant, only that prompt specificity did not appear to explain students' judgments in this activity.

The contrast in students' validation judgments depended largely on whether they could use physics and mathematics to examine the reasoning in the AI-generated solution. For students who had solved the problem at least partly, comparing the generated response with their own attempts helped turn their initial criteria into specific checks on the solution, rather than on its appearance or final answer alone. In these cases, validation was possible because students could connect parts of the AI-generated reasoning with the physical situation, the mathematical development, and the criteria they had stated before seeing the problem or the AI-generated response.

For students who struggled with the problem, comparison with the AI-generated solution played a diagnostic role. Because the task required setting up and interpreting an integral, difficulty following the mathematical development limited their ability to judge the validity of the response. For them, the comparison was less useful for validating the response than for making visible where understanding broke down: in defining the differential charge element, choosing variables, carrying out the algebra, or interpreting components. In these cases, students' responses highlighted what they needed to understand before they could judge the solution.

Taken together, these responses suggest that students should be explicitly taught how to validate solutions as part of problem solving itself. In this framing, validation is not a final check applied after an answer is produced; it is something students learn to do while examining how a solution is constructed, represented, and evaluated. This requires structured opportunities to practice problem-solving moves as tools for validation: identifying the model being used, checking whether the relevant information has been used appropriately, making assumptions explicit, following the logical progression of the reasoning, checking units and limiting cases, and evaluating whether both the result and the assumptions remain reasonable.[7]

This kind of validation work can also reshape how students use AI-generated solutions. Because AI can now produce answers to many routine problems, moving some work into supervised classroom settings may be necessary for certain instructional goals.[8] But another response is to treat AI-generated solutions as objects for analysis. When students ask AI to solve a problem, the generated response does not have to function only as an answer to be accepted or rejected; it can also be treated as a worked-example-like solution. Research on worked examples has shown that, under some conditions, studying worked-out solutions can support later problem solving, but the benefit depends on how those examples are structured and how students are guided to use them.[9] In that setting, the question is not simply whether students use AI-generated solutions, but whether they know how to study them critically: reconstructing the reasoning, noticing omissions and making assumptions explicit, asking follow-up questions that clarify the solution, and checking whether the argument is physically and mathematically justified. For students who cannot yet follow the mathematical reasoning, this work may begin by locating where understanding breaks down. Still, asking the tool to explain a part of the solution is not the same as understanding it. The value of the AI response lies not in providing a finished answer, but in giving students a worked-out response they can reconstruct, question, and test.

Given the characteristics of this study, these results are not generalizable to all educational contexts. The mathematical demands of the problem may have contributed to students' caution and should also be considered when interpreting the findings. With a simpler task, for which students could more readily follow the reasoning and check the result, their first validation judgments might differ, including a greater willingness to accept an AI-generated solution after a first reading. Because the activity was structured, the results do not describe students' spontaneous use of AI. Under these conditions, students did not treat the AI-generated solution as valid automatically or uniformly. Instead, their judgments depended on what they could examine in the response. This suggests that teaching students to use AI critically requires more than warning them that the tool can be wrong.

## 5. References

1. Krupp, L., Steinert, S., Kiefer-Emmanouilidis, M., Avila, K. E., Lukowicz, P., Kuhn, J., ... & Karolus, J. (2023). Unreflected acceptance - Investigating the Negative Consequences of ChatGPT-Assisted Problem Solving in Physics Education. *arXiv preprint arXiv:2309.03087*.
2. Avila, K. E., Steinert, S., Ruzika, S., Kuhn, J., & Küchemann, S. (2024). Using ChatGPT for teaching physics. *The Physics Teacher*, *62*(6), 536-537.
3. Suárez, Á. (2025). Teaching with Questions, Not Answers: ChatGPT as a Socratic Partner in Physics Learning. *The Physics Teacher*, *63*(9), 804-805.
4. Wu, H., Tong, D., Li, Z., Tao, Y., Deng, S., & Ren, H. (2025). GPT's Application in Physics Education: A Rapid Review. *The Physics Teacher*, *63*(9), 786-790.
5. Dahlkemper, M. N., Lahme, S. Z., & Klein, P. (2023). How do physics students evaluate artificial intelligence responses on comprehension questions? A study on the perceived scientific accuracy and linguistic quality of ChatGPT. *Physical Review Physics Education Research*, *19*(1), 010142.
6. Sadidi, F., & Prestel, T. (2025). Impact of criterion-based reflection on prospective physics teachers' perceptions of ChatGPT-generated content. *Discover Education*, *5*(1), 57.
7. Burkholder, E. W., Miles, J. K., Layden, T. J., Wang, K. D., Fritz, A. V., & Wieman, C. E. (2020). Template for teaching and assessment of problem solving in introductory physics. *Physical Review Physics Education Research*, *16*(1), 010123.
8. Kortemeyer, G. (2026). The boiling-frog problem of physics education. *The Physics Teacher*, *64*(1), 8-12.
9. Atkinson, R. K., Derry, S. J., Renkl, A., & Wortham, D. (2000). Learning from examples: Instructional principles from the worked examples research. *Review of Educational Research*, *70*(2), 181-214.

## Appendix A. Student activity

### Phase 1 – Individual work

Suppose a classmate solves a physics problem and gives you their solution to look over. Before solving anything, think about that situation and answer the following question:

What three features would you look at to decide whether that solution is valid?

Each feature should refer to a part or characteristic of the solution clearly enough that another person could understand what you mean. Avoid answers that are too general.

Write your response here:

### Phase 2 – Small-group work

In groups of two or three, solve the following problem on the small boards. You do not need to obtain the full exact solution, but you should attempt at least part (a) and one of the two limiting cases in part (b) or part (c). Once your response is ready on the board, let the instructor know so that the responses can be recorded.

A uniformly charged rod of length $L$, total charge $Q$, and linear charge density $\lambda$ lies along the x-axis, as shown in the figure. Consider point P, located on the y-axis at $(0, d)$.

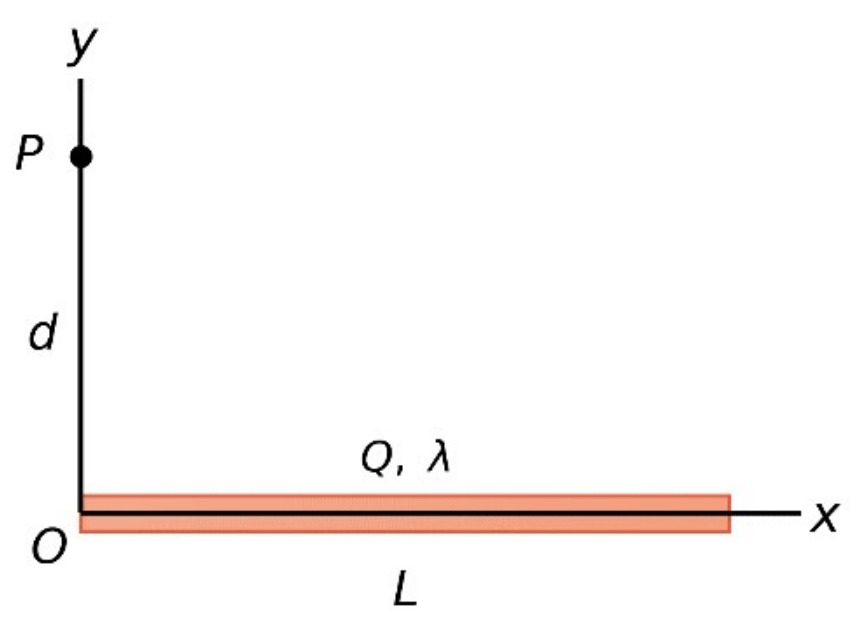

a) Determine the components $E_x$ and $E_y$ of the electric field at point P.
b) Consider the limit $d \gg L$. If you obtained the exact expression in part (a), use it to obtain an approximate expression for $E_y$ in this limit. If you did not obtain the exact expression, identify the simpler physical system with which the rod can be compared in this case and, based on that comparison, determine an approximate expression for $E_y$.
c) Consider the limit $L \gg d$. If you obtained the exact expression in part (a), use it to obtain an approximate expression for $E_y$ in this limit. If you did not obtain the exact expression, identify the simpler charge distribution with which the rod can be compared in this case and, based on that comparison, determine how $E_y$ should depend on $d$ and what approximate value you would expect for that component.

### Phase 3 – Individual work

Now solve this problem using ChatGPT, Gemini, or another AI you're familiar with.

1. Upload to ChatGPT or Gemini a screenshot of the problem statement exactly as it appears in the previous part. Ask the tool to solve the problem in whatever way you find most useful.
2. Read the full solution once, from beginning to end, and do not interact further with the chat.
3. Share the link to the conversation in the worksheet.

Paste your link here:

4. After that first reading, and before discussing the solution with your group or returning to it, answer the following questions individually.

### *3.1 Initial decision*

Highlight the option you consider most appropriate:

a) After this first reading, I would accept the solution as valid right away.
b) After this first reading, I lean toward considering it valid, but before accepting it as valid I would need to check one or two specific aspects.
c) After this first reading, I would not accept it as valid.

### *3.2 Justification of that decision*

a) What I focused on most when making that decision was:
   Write your response here:
b) Only if you chose B or C: What aspects would you need to check before accepting it as valid?
   Write what you think here:

## Phase 4 – Small-group discussion and whole-class exchange

Return to your original small groups. Share and compare your individual responses. Use the following questions to guide your discussion.

a) When you compare your individual responses, what similarities and differences do you notice in what each group member said they focused on when making their initial decision?
b) Read again the three features each group member wrote in Phase 1. How are they similar to, and how are they different from, what you said you focused on when reading the AI-generated solution?
c) If you had to explain to another group what this activity helped you notice about evaluating an AI-generated solution, what would you say?

After the small-group discussion, the class will hold a brief whole-class exchange to share the main points that emerged.

## Appendix B. Examples of AI-generated solutions

This appendix includes two complete AI-generated solutions accessed through the conversation links provided by the students. The responses are shown in Spanish, as generated in the original conversations. The examples were selected to represent the two ends of the observed variation in response detail: the most detailed and the most concise response in the set.

**First solution:** ChatGPT (GPT-5.4 mini), most detailed response
**Second solution:** Gemini (Gemini 3 Flash), most concise response

Esta es una copia de un chat compartido de ChatGPT. No se añadirá a la memoria.
**Informar sobre la conversación**

**lun, 20 abr** a la(s) 23:08

Una varilla uniformemente cargada de longitud $L$, carga total $Q$ y densidad lineal ¿$Q/L$, se encuentra ubicada a lo largo del eje $x$, como se muestra en la figura. Consideren el punto P, ubicado sobre el eje $y$ en $(0, d)$.

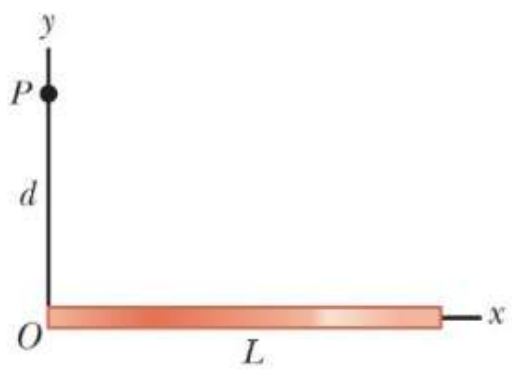

a) Determinen las componentes $E_x$ y $E_y$ del campo eléctrico en el punto P.
b) Analicen el caso $d \gg L$. Si obtuvieron la expresión exacta en (a), utilícenla para hallar una expresión aproximada de $E_y$ en ese régimen. Si no la obtuvieron, indiquen con qué sistema físico más simple puede compararse la varilla en ese caso y, a partir de esa comparación, determinen una expresión aproximada de $E_y$.
c) Analicen el caso $L \gg d$ Si obtuvieron la expresión exacta en (a), utilícenla para hallar una expresión aproximada de $E_y$ en ese régimen. Si no la obtuvieron, indiquen con qué distribución de carga más simple puede compararse la varilla en ese caso y, a partir de esa comparación, determinen cómo debería depender $E_y$ de $d$ y qué valor aproximado esperarían para esa componente.

Archivo cargado

quiero que me resuelvas este ejercicio paso a paso utilizando este libro como referencia, el nivel es terciaro (estudiante de profesorado de física)
indica los supuestos iniciales, y en el paso a paso explica los porqués físicos aplicando las maneras del Serway para poder comprender cada decisión.

Pensado durante 36s >

Sí. Lo resolvemos con el enfoque "a la Serway": **conceptualizar, categorizar, analizar y finalizar**, usando superposición e integración para una distribución continua de carga. El problema coincide con la temática de **campo eléctrico de una distribución continua de carga** del capítulo 23, y el propio libro recomienda justamente dividir la carga en elementos infinitesimales $dq$, sumar vectorialmente sus aportes y aprovechar la simetría del sistema. Serway+y+Jewett+10ed+vol+2 Además, en la sección de problemas aparece este mismo ejercicio de la varilla sobre el eje $x$ y el punto $P$ sobre el eje $y$. Serway+y+Jewett+10ed+vol+2

## 1) Supuestos iniciales

Antes de calcular, dejo explícitos los supuestos físicos:

1. **Distribución lineal uniforme de carga**: la varilla tiene carga total $Q$ repartida uniformemente a lo largo de una longitud $L$, de modo que

$$\lambda = \frac{Q}{L}.$$

2. **Régimen electrostático**: las cargas están fijas, no hay corrientes ni dependencia temporal.
3. **Medio homogéneo**: asumimos vacío o aire idealizado, de modo que vale la ley de Coulomb con

$$k_e = \frac{1}{4\pi\varepsilon_0}.$$

4. **Varilla delgada**: la modelamos como una línea de carga, despreciando su espesor transversal.
5. **Punto de observación**: $P = (0, d)$, con $d > 0$, mientras la varilla ocupa el tramo del eje $x$ desde $x = 0$ hasta $x = L$. Eso es importante: la varilla **no está centrada** respecto del eje $y$, por eso no se anula la componente horizontal.

---

## 2) Conceptualizar

Queremos el campo eléctrico total en $P$.

La idea física central es esta: cada pedacito de varilla produce en $P$ un pequeño campo $d\vec{E}$, y el campo total es la suma vectorial de todos esos aportes. Eso es exactamente el principio de superposición aplicado a una distribución continua. Serway+y+Jewett+10ed+vol+2

Ahora pensemos la dirección:

- si $Q > 0$, cada $d\vec{E}$ en $P$ apunta **alejándose** del elemento de carga;
- como todos los elementos están a la derecha del eje $y$, cada aporte tendrá:
  - una componente en $x$ hacia la **izquierda** $(-\hat{i})$,
  - una componente en $y$ hacia **arriba** $(+\hat{j})$.

Entonces, antes de calcular ya esperamos:

$$E_x < 0, \qquad E_y > 0$$

si $Q > 0$.
Si $Q < 0$, ambas componentes cambian de signo.

---

## 3) Categorizar

No es una suma de cargas puntuales aisladas, sino una **distribución continua lineal**. Entonces, como sugiere Serway, pasamos de suma a integral y escribimos $dq = \lambda\, dx$.

Serway+y+Jewett+10ed+vol+2

---

## 4) Analizar: planteo diferencial

Tomemos un elemento de carga en la posición $x$ de la varilla.

Entonces:

- su carga es

$$dq = \lambda\, dx$$

- el vector que va desde ese elemento $(x, 0)$ hasta el punto $P = (0, d)$ es

$$\vec{r} = (-x)\,\hat{\imath} + d\,\hat{\jmath}$$

- su módulo es

$$r = \sqrt{x^2 + d^2}$$

Por ley de Coulomb, el módulo del campo elemental es

$$dE = k_e \frac{dq}{r^2} = k_e \frac{\lambda\, dx}{x^2 + d^2}.$$

Pero no queremos solo el módulo: necesitamos componentes.
Para eso usamos la dirección del vector $\vec{r}$. Su vector unitario es

$$\hat{r} = \frac{-x\,\hat{\imath} + d\,\hat{\jmath}}{\sqrt{x^2 + d^2}}.$$

Entonces

$$d\vec{E} = k_e \frac{dq}{r^2}\hat{r} = k_e \frac{\lambda\, dx}{x^2 + d^2} \cdot \frac{-x\,\hat{\imath} + d\,\hat{\jmath}}{\sqrt{x^2 + d^2}}.$$

Queda

$$d\vec{E} = k_e \lambda \frac{-x\,\hat{\imath} + d\,\hat{\jmath}}{(x^2 + d^2)^{3/2}}\, dx.$$

De ahí salen directamente las componentes:

$$dE_x = -\,k_e \lambda \frac{x}{(x^2 + d^2)^{3/2}}\, dx$$

$$dE_y = k_e \lambda \frac{d}{(x^2 + d^2)^{3/2}}\, dx$$

Esto ya muestra algo físico importante:

- el signo menos de $dE_x$ aparece porque todos los aportes empujan hacia la izquierda;

- $dE_y$ es positivo porque todos empujan hacia arriba.

---

## 5) Integración de $E_x$

Integramos desde $x = 0$ hasta $x = L$:

$$E_x = \int_0^L dE_x = -k_e\lambda \int_0^L \frac{x}{(x^2 + d^2)^{3/2}}\, dx.$$

Hacemos el cambio

$$u = x^2 + d^2, \qquad du = 2x\, dx.$$

Entonces

$$\int \frac{x}{(x^2 + d^2)^{3/2}}\, dx = \frac{1}{2}\int u^{-3/2}\, du = -\frac{1}{\sqrt{u}} = -\frac{1}{\sqrt{x^2 + d^2}}.$$

Por tanto,

$$E_x = -k_e\lambda \left[-\frac{1}{\sqrt{x^2 + d^2}}\right]_0^L.$$

Evaluando:

$$E_x = -k_e\lambda \left(-\frac{1}{\sqrt{L^2 + d^2}} + \frac{1}{d}\right).$$

Ordenando,

$$\boxed{E_x = k_e\lambda \left(\frac{1}{\sqrt{L^2 + d^2}} - \frac{1}{d}\right)}$$

Como $\sqrt{L^2 + d^2} > d$, se cumple

$$\frac{1}{\sqrt{L^2 + d^2}} < \frac{1}{d},$$

así que efectivamente $E_x < 0$, tal como esperábamos físicamente.

También puede escribirse como

$$\boxed{E_x = -\, k_e\lambda \left(\frac{1}{d} - \frac{1}{\sqrt{L^2 + d^2}}\right)}$$

---

## 6) Integración de $E_y$

Ahora:

$$E_y = \int_0^L dE_y = k_e \lambda d \int_0^L \frac{dx}{(x^2 + d^2)^{3/2}}.$$

Usamos la integral conocida

$$\int \frac{dx}{(x^2 + a^2)^{3/2}} = \frac{x}{a^2\sqrt{x^2 + a^2}}.$$

Con $a = d$,

$$E_y = k_e \lambda d \left[ \frac{x}{d^2\sqrt{x^2 + d^2}} \right]_0^L.$$

Entonces:

$$E_y = k_e \lambda d \cdot \frac{L}{d^2\sqrt{L^2 + d^2}}$$

y queda

$$\boxed{E_y = k_e \lambda \frac{L}{d\sqrt{L^2 + d^2}}}$$

Usando $\lambda = Q/L$,

$$\boxed{E_y = k_e \frac{Q}{d\sqrt{L^2 + d^2}}}$$

y también

$$\boxed{E_x = k_e \frac{Q}{L} \left( \frac{1}{\sqrt{L^2 + d^2}} - \frac{1}{d} \right)}$$

---

## 7) Resultado vectorial exacto del inciso (a)

Entonces el campo eléctrico en $P$ es

$$\boxed{\vec{E}(P) = k_e \lambda \left( \frac{1}{\sqrt{L^2 + d^2}} - \frac{1}{d} \right) \hat{\imath} + k_e \lambda \frac{L}{d\sqrt{L^2 + d^2}} \hat{\jmath}}$$

o, en función de $Q$,

$$\boxed{\vec{E}(P) = k_e \frac{Q}{L}\left(\frac{1}{\sqrt{L^2+d^2}} - \frac{1}{d}\right)\hat{i} + k_e \frac{Q}{d\sqrt{L^2+d^2}}\hat{j}}$$

Para $Q > 0$:

- $E_x < 0$,
- $E_y > 0$.

---

## 8) Inciso (b): caso $d \gg L$

### Idea física primero

Si el punto está muy lejos comparado con la longitud de la varilla, la distribución “se ve” casi como una **carga puntual** $Q$ ubicada cerca del origen. Esa es exactamente la lógica física que Serway usa repetidamente: a grandes distancias, el detalle geométrico de la distribución deja de importar y domina la carga total. En problemas del capítulo también se pide comparar distribuciones extendidas con partículas cuando la distancia es mucho mayor que el tamaño del sistema.

### Aproximación de $E_y$

Partimos de la expresión exacta:

$$E_y = k_e \frac{Q}{d\sqrt{L^2+d^2}}.$$

Si $d \gg L$,

$$\sqrt{L^2+d^2} = d\sqrt{1+\frac{L^2}{d^2}} \approx d.$$

Entonces

$$\boxed{E_y \approx k_e \frac{Q}{d^2}}$$

Esto coincide con el campo de una carga puntual $Q$ a distancia $d$.

### ¿Y qué pasa con $E_x$?

De la exacta:

$$E_x = k_e \frac{Q}{L}\left(\frac{1}{\sqrt{L^2+d^2}} - \frac{1}{d}\right).$$

Si $d \gg L$, desarrollando un poco:

$$\frac{1}{\sqrt{L^2+d^2}} = \frac{1}{d}\left(1+\frac{L^2}{d^2}\right)^{-1/2} \approx \frac{1}{d}\left(1-\frac{L^2}{2d^2}\right).$$

Luego

$$E_x \approx k_e \frac{Q}{L}\left(\frac{1}{d} - \frac{L^2}{2d^3} - \frac{1}{d}\right) = -\, k_e \frac{QL}{2d^3}.$$

O sea,

$$\boxed{E_x \approx -\, k_e \frac{QL}{2d^3}}$$

Esto muestra que $E_x$ cae más rápido que $E_y$, por eso a gran distancia el campo queda casi vertical.

---

## 9) Inciso (c): caso $L \gg d$

### Idea física primero

Si la varilla es muchísimo más larga que la distancia al punto $P$, el punto "ve" a la varilla casi como una **línea semiinfinita** que parte del origen y se extiende hacia $+x$. No puede compararse con una línea infinita completa, porque esa tendría simetría arriba-abajo respecto al punto medio y aquí no la hay: la varilla empieza en $x = 0$, no sigue hacia $x < 0$.

### Aproximación de $E_y$

Tomamos la exacta:

$$E_y = k_e \frac{Q}{d\sqrt{L^2+d^2}}$$

o mejor, usando $\lambda$,

$$E_y = k_e \lambda \frac{L}{d\sqrt{L^2+d^2}}.$$

Si $L \gg d$,

$$\sqrt{L^2+d^2} \approx L$$

y entonces

$$\boxed{E_y \approx k_e \frac{\lambda}{d}}$$

Como $k_e = \dfrac{1}{4\pi\varepsilon_0}$,

$$\boxed{E_y \approx \frac{\lambda}{4\pi\varepsilon_0\, d}}$$

### Interpretación física

Este resultado tiene mucho sentido:

- depende de la **densidad lineal** $\lambda$, no de la carga total $Q$, porque si la varilla es muy larga el extremo lejano aporta muy poco;
- varía como $1/d$, que es la dependencia típica de campos asociados a distribuciones lineales extensas.

### Valor esperado en comparación con una línea infinita

Para una línea **infinita**,

$$E_\infty = \frac{\lambda}{2\pi\varepsilon_0\, d} = 2k_e \frac{\lambda}{d}.$$

Nuestro valor es

$$E_y \approx k_e \frac{\lambda}{d} = \frac{1}{2} E_\infty.$$

Eso es perfectamente razonable: una varilla semiinfinita produce, sobre este punto, aproximadamente **la mitad** del campo vertical que una línea infinita, porque "falta" toda la mitad situada hacia $x < 0$.

### De hecho, también $E_x$ se aproxima

De la exacta:

$$E_x = k_e \lambda \left( \frac{1}{\sqrt{L^2 + d^2}} - \frac{1}{d} \right)$$

si $L \gg d$, entonces $1/\sqrt{L^2 + d^2} \approx 1/L \approx 0$, y queda

$$\boxed{E_x \approx -k_e \frac{\lambda}{d}}$$

Así que, en este régimen,

$$\boxed{\vec{E} \approx k_e \frac{\lambda}{d} \left( -\hat{i} + \hat{j} \right)}$$

y su módulo sería

$$|\vec{E}| \approx \sqrt{2}\, k_e \frac{\lambda}{d}.$$

Esto además muestra algo lindo físicamente: para la semiinfinita, cuando $L \gg d$, las componentes quedan aproximadamente iguales en módulo.

---

## 10) Resumen final ordenado

### (a) Componentes exactas

$$\boxed{E_x = k_e\lambda\left(\frac{1}{\sqrt{L^2+d^2}} - \frac{1}{d}\right)}$$

$$\boxed{E_y = k_e\lambda\frac{L}{d\sqrt{L^2+d^2}}}$$

o, con $\lambda = Q/L$,

$$\boxed{E_x = k_e\frac{Q}{L}\left(\frac{1}{\sqrt{L^2+d^2}} - \frac{1}{d}\right)}$$

$$\boxed{E_y = k_e\frac{Q}{d\sqrt{L^2+d^2}}}$$

### (b) Si $d \gg L$

La varilla se comporta como una **carga puntual** $Q$:

$$\boxed{E_y \approx k_e\frac{Q}{d^2}}$$

y además

$$\boxed{E_x \approx -\,k_e\frac{QL}{2d^3}}$$

### (c) Si $L \gg d$

La varilla se comporta como una **línea semiinfinita**:

$$\boxed{E_y \approx k_e\frac{\lambda}{d} = \frac{\lambda}{4\pi\varepsilon_0 d}}$$

y también

$$\boxed{E_x \approx -k_e \frac{\lambda}{d}}$$

---

## 11) Lectura física final, bien “Serway”

Lo más importante para comprender la decisión matemática es esto:

- **No empezamos integrando porque sí**.
- Primero identificamos el **modelo físico**: distribución continua lineal.
- Luego usamos el **principio fundamental**: superposición.
- Después elegimos una partición elemental natural del sistema: $dq = \lambda dx$.
- Finalmente aprovechamos la **geometría** para separar componentes y sumar sólo lo que corresponde.

Ese es exactamente el tipo de razonamiento que el libro enfatiza: no memorizar fórmulas aisladas, sino reconocer el sistema, elegir el modelo correcto y construir el resultado desde el principio físico adecuado. Serway+y+Jewett+10ed+vol+2

Si querés, en el siguiente mensaje te lo puedo rehacer en formato “**para entregar**”, más prolijo y corrido, como desarrollo formal de profesorado.

Fuentes

Fuentes

de longitud $L$ y carga tota
nuestra en la figura.
npo eléctrico en el punto $P$ ut
desde el origen.
le la componente vertical del c
ica por qué se podía esperar

e la componente vertical del ca
sultado.

resolve este ejercico explicando el paso a paso

Para resolver este problema de electrostática, utilizaremos el principio de superposición integrando los campos eléctricos generados por cada pequeño elemento de carga de la varilla.

## Configuración inicial

Definamos primero los parámetros del problema:

- La varilla tiene longitud $L$ y carga total $Q$.
- La densidad lineal de carga es $\lambda = \frac{Q}{L}$.
- Consideramos un elemento diferencial de carga $dq$ situado en una posición $x$ sobre el eje $x$. Por tanto, $dq = \lambda dx = \frac{Q}{L}dx$.
- El punto $P$ está en la coordenada $(0, d)$.
- La distancia desde $dq$ al punto $P$ es $r = \sqrt{x^2 + d^2}$.

El campo eléctrico diferencial $dE$ producido por $dq$ en el punto $P$ tiene una magnitud:

$$dE = \frac{k\,dq}{r^2} = \frac{k(\frac{Q}{L}dx)}{x^2 + d^2}$$

Donde $k = \frac{1}{4\pi\epsilon_0}$ es la constante de Coulomb.

## a) Componentes del campo eléctrico en el punto $P$

Descomponemos $dE$ en sus componentes $x$ e $y$ basándonos en el ángulo $\theta$ que forma el vector con el eje $y$. Observando la geometría:

$\cos\theta = \frac{d}{r} = \frac{d}{\sqrt{x^2+d^2}}$ (componente vertical)

$\sin\theta = \frac{x}{r} = \frac{x}{\sqrt{x^2+d^2}}$ (componente horizontal, hacia la izquierda, por eso negativa)

**Componente vertical ($E_y$):**

$$E_y = \int dE_y = \int_0^L \frac{k\,dq}{r^2}\cos\theta = \int_0^L \frac{k\frac{Q}{L}dx}{x^2+d^2}\cdot\frac{d}{\sqrt{x^2+d^2}}$$

$$E_y = \frac{kQd}{L}\int_0^L \frac{dx}{(x^2+d^2)^{3/2}}$$

Usando la integral estándar $\int \frac{dx}{(x^2+a^2)^{3/2}} = \frac{x}{a^2\sqrt{x^2+a^2}}$:

$$E_y = \frac{kQd}{L}\left[\frac{x}{d^2\sqrt{x^2+d^2}}\right]_0^L = \frac{kQd}{L}\left(\frac{L}{d^2\sqrt{L^2+d^2}}\right)$$

$$E_y = \frac{kQ}{d\sqrt{L^2+d^2}}$$

**Componente horizontal ($E_x$):**

$$E_x = \int dE_x = \int_0^L \frac{k\,dq}{r^2}(-\sin\theta) = -\frac{kQ}{L}\int_0^L \frac{x\,dx}{(x^2+d^2)^{3/2}}$$

Haciendo un cambio de variable $u = x^2 + d^2$, $du = 2xdx$:

$$E_x = -\frac{kQ}{L}\left[-\frac{1}{\sqrt{x^2+d^2}}\right]_0^L = -\frac{kQ}{L}\left(\frac{1}{d} - \frac{1}{\sqrt{L^2+d^2}}\right)$$

## b) Caso $d \gg L$

Cuando la distancia $d$ es mucho mayor que la longitud de la varilla ($d \gg L$), podemos aproximar el comportamiento del campo vertical $E_y$:

$$E_y = \frac{kQ}{d\sqrt{L^2 + d^2}} \approx \frac{kQ}{d\sqrt{d^2}} = \frac{kQ}{d^2}$$

**Explicación:**
Este resultado es idéntico al campo eléctrico creado por una **carga puntual**. Esto es de esperar porque, desde una distancia muy grande, la varilla es tan pequeña en comparación con la distancia al observador que toda la carga parece concentrada en un solo punto (el origen).

## c) Caso $L \gg d$

Cuando la longitud de la varilla es mucho mayor que la distancia $d$ ($L \gg d$), el término $\sqrt{L^2 + d^2} \approx \sqrt{L^2} = L$. Entonces:

$$E_y = \frac{kQ}{d\sqrt{L^2 + d^2}} \approx \frac{kQ}{dL} = \frac{k(Q/L)}{d} = \frac{k\lambda}{d}$$

**Explicación:**
En este límite, la varilla se comporta como una **línea de carga semi-infinita** vista desde su extremo. El campo decae con $1/d$ (linealmente con la distancia) en lugar de $1/d^2$ (como una carga puntual). Esto ocurre porque, al ser tan larga, la contribución de la carga se distribuye a lo largo de una línea infinita, lo cual es una característica típica de los campos producidos por distribuciones de carga 1D infinitas.

¿Te gustaría profundizar en cómo cambia el campo si el punto $P$ estuviera centrado respecto a la varilla en lugar de estar en el extremo?